\documentclass[fleqn,twoside]{article}
\usepackage{espcrc2}

\usepackage{graphicx, psfig}
\usepackage[figuresright]{rotating}

\newcommand{\AmS}{{\protect\the\textfont2
  A\kern-.1667em\lower.5ex\hbox{M}\kern-.125emS}}

\title{The Ultra Luminous Infrared Galaxy MKN~231: \\New clues from {\it Beppo}SAX  and XMM-{\it Newton} }

\author{V. Braito\address[Brera]{INAF $-$ Osservatorio Astronomico di Brera, Via Brera 28, 20154 Milano,
Italy}\address[Padova]{Dipartimento di Astronomia, Universit\`a di  Padova, Vicolo Dell'Osservatorio 2, 35122 Padova, Italy}\thanks{This work
has received financial support from ASI (I/R/037/01)  and from ASI (I/R/062/02) under the project ``Cosmologia Osservativa con
XMM-Newton".}, R. Della Ceca\addressmark[Brera], E. Piconcelli\address[cnrbo]{IASF $-$ CNR, Sezione di Bologna, Via Gobetti 101, 40129 Bologna,
Italy}\address[vilspa]{XMM-SOC (VILSPA),  ESA,  28080 Madrid, Spain }, P. Severgnini\addressmark[Brera], L.\,Bassani\addressmark[cnrbo], 
M.\,Cappi\addressmark[cnrbo], A.\,Franceschini\addressmark[Padova], K. Iwasawa\address[]{Institute of Astronomy, University of Cambridge,
Madingley Road Cambridge CB3 OHA, U.K. }, G.\,Malaguti\addressmark[cnrbo], P. Marziani\address[]{INAF $-$ Osservatorio Astronomico di
Padova, Vicolo Dell'Osservatorio 5, 35122 Padova, Italy}, G.G.C.\,Palumbo\address[]{Dipartimento di Astronomia, Universit\`a di Bologna, Via
Ranzani 1, 40127 Bologna, Italy}, M.\,Persic\address[]{INAF $-$ Osservatorio Astronomico di Trieste,  Via Tiepolo 11, 34131 Trieste, 
Italy}, G.\,Risaliti\address[arc]{INAF $-$ Osservatorio Astrofisico di Arcetri, L.go E. Fermi  5, 50125  Firenze,
Italy}\address[]{Harvard-Smithsonian Center for Astrophysics, Cambridge, USA} and  M.\,Salvati\addressmark[arc]}

\begin{document}

\begin{abstract}  
We present {\it Beppo}SAX and XMM-{\it Newton}  observations of MKN 231. These observations  and in particular the BeppoSAX PDS
data allowed us to unveil, for the first time, the highly absorbed (N$_H\sim  2\times10^{24}$ cm$^{-2}$) AGN
component. We find that: a) the AGN powering MKN231 has an intrinsic 2-10 keV luminosity of
$1^{+1}_{-0.5}\times
10^{44}$ erg/s; b) the strong
starburst activity   contributes significantly in the 0.1-10 keV energy range. We propose that the 
starburst activity strongly contributes to the  far infrared luminosity of MKN 231;  this is also suggested by the
multiwavelength properties of MKN 231.

\vspace{1pc}
\end{abstract}

\maketitle

\section{INTRODUCTION} 

MKN~231 is one of the best studied ULIRG and one of the most luminous object in the local universe.
  Although  the nature of the primary energy
source of ULIRGs (AGN vs. starburst activity)   still remains rather enigmatic
(see \cite{France} and reference therein),  in the case of   MKN~231 optical and near infrared observations seem to suggest that a
significant contribution to the infrared luminosity could be ascribed to AGN
activity \cite{golda,krab}.  Moreover  MKN~231  is also classified as a Broad
Absorption line (BAL) QSO \cite{Smith}. On the other hand this
galaxy is also undergoing an energetic starburst. This  strong starburst
activity  combined with  the luminous  AGN   makes MKN~231 one of the best
example of the transition from starburst  to AGN according to the scheme
outlined in   \cite{san}. The presence of absorption possibly associated with
the BAL outflows combined with the   strong starburst activity, makes the X-ray
continuum of MKN~231 particularly complex.  ROSAT, ASCA and {\it Chandra}
(\cite{galla,iwa,malo,turn}) observations have shown this complexity revealing 
both an  extended soft X-ray emission (associated to the starburst activity)
and a hard  X-ray emission indicative of a heavily obscured AGN. The models
proposed to explain the flat X-ray spectra of MKN~231 accumulated so far
(which were limited to
E$<$10 keV), invoke  reprocessed emission: thus the  estimate of the
intrinsic power of the AGN can be obtained only through indirect arguments.

\section{Observations and data reduction}
MKN~231 was targeted by XMM-{\it Newton}   on July 6th, 2001  and by \emph{Beppo}SAX  \cite{boella} from December
29th 2001 to January 1st 2002 (see Table 1). For  the scientific analysis of the  \emph{Beppo}SAX observation only the data  collected from
the MECS  and from  the PDS have been considered. The XMM-{\it Newton} EPIC data  have been  filtered  from high background time intervals and only events corresponding to
pattern 0-12 for MOS and pattern 0-4 for pn have been used.
\begin{table}[thb]
\caption{Exposure and count rates}
\label{table:1}
\newcommand{\m}{\hphantom{$-$}}
\newcommand{\cc}[1]{\multicolumn{1}{c}{#1}}
\renewcommand{\tabcolsep}{1pc} % enlarge column spacing
\renewcommand{\arraystretch}{1.2} % enlarge line spacing
\begin{tabular}{@{}ccc}
\hline
Instrument           & EXPOSURE & Cts/s  \\
& ks& $10^{-2}$ \\
\hline
  pn			& 16	&	$11\pm1	     $	\\
  MOS2		& 19.8	&	$3.54\pm 0.14$	\\
  MOS1		& 19.9	&	$3.77\pm 0.14$	\\
PDS     		& 76  	&	$6.58\pm2.16 $	\\
MECS    		& 144 	&	$0.57\pm0.02 $	\\
 
\hline
\end{tabular}
\end{table}
 \begin{figure*}[thb]
%\vspace{9pt}
{\psfig{file=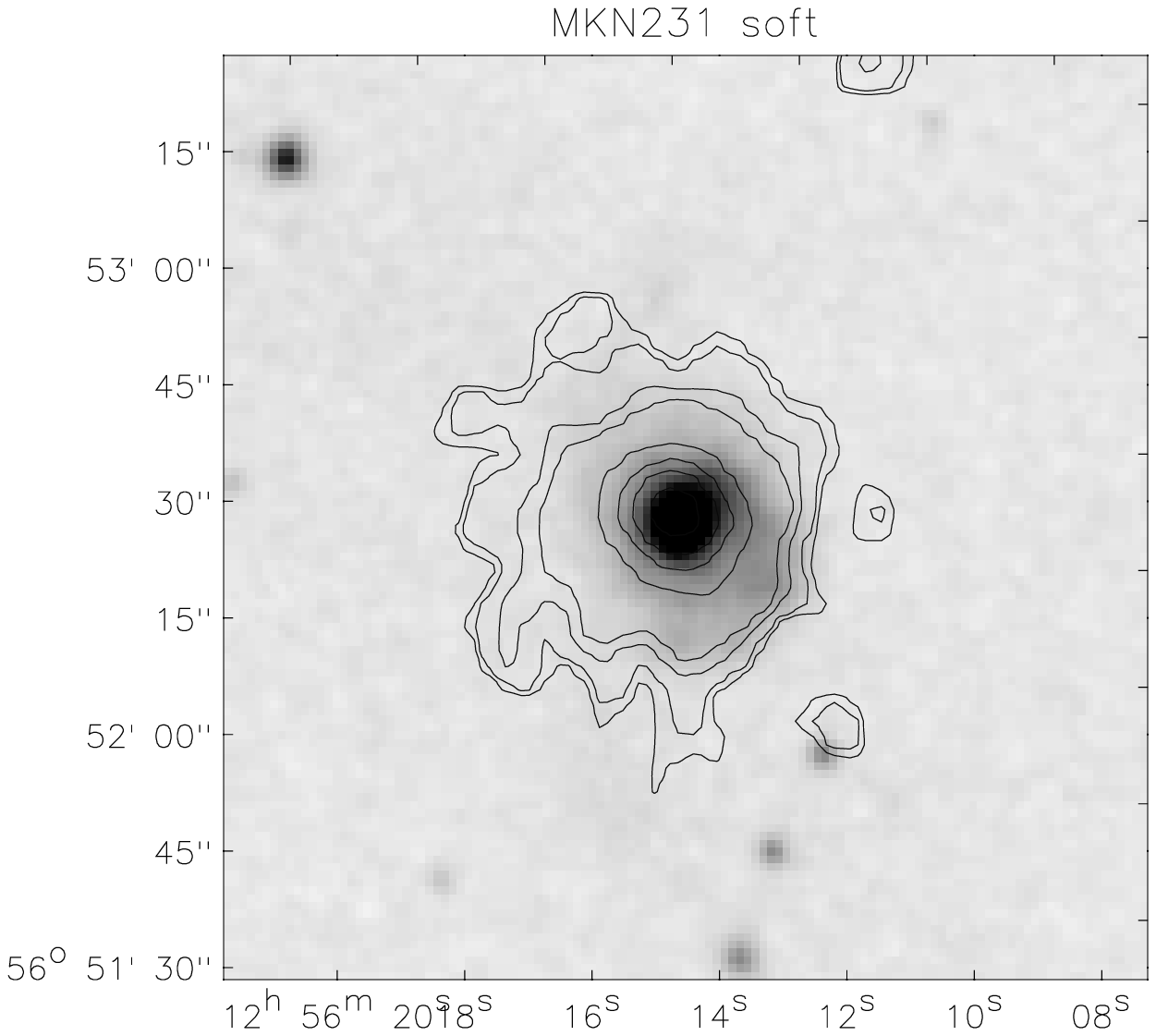,width=0.6\textwidth } \vskip-6.8truecm \hskip7.truecm   \psfig{file=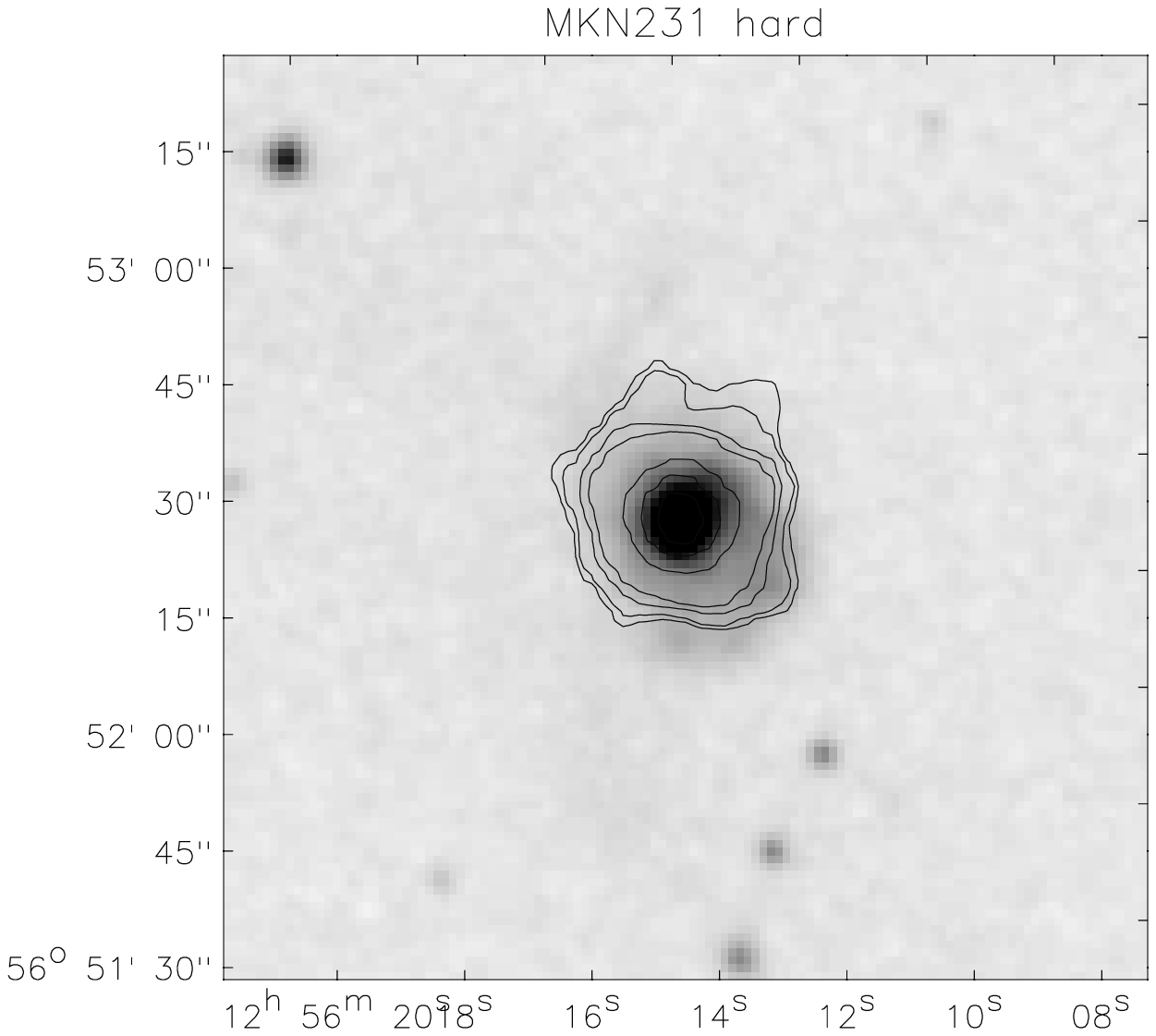,width=0.6\textwidth}}
\skip-0.5truecm
\caption{DSS2 image ($2'\times 2'$) of the field centered  on MKN~231.   Contours of the soft (0.5$-$2 keV; left panel ) and hard (4.5--10 keV; right panel)  
X-ray emission  have been   overlaid on the optical image.  The contours 
displayed  correspond to 3$\sigma$, 4$\sigma$, 7$\sigma$, 10$\sigma$, 30$\sigma$, 
50$\sigma$  above the background. }
\label{fig1}
 \end{figure*}
  In Fig.~\ref{fig1} the XMM-{\it Newton}  0.5--2 keV (left
panel) and the 4.5--10 keV (right panel)   X-ray contours  are overlaid to the optical DSS2-red image of MKN~231.  The X-ray
emission of MKN 231 appears to be extended  in the low energy domain ($E<$ 2 keV), while the  hard (4.5--10.0 keV) X-ray
brightness profile is comparable to the XMM-{\it Newton} point spread function (PSF). 

\section{The 0.5--50 keV  spectrum.}

\begin{figure*}[thb]
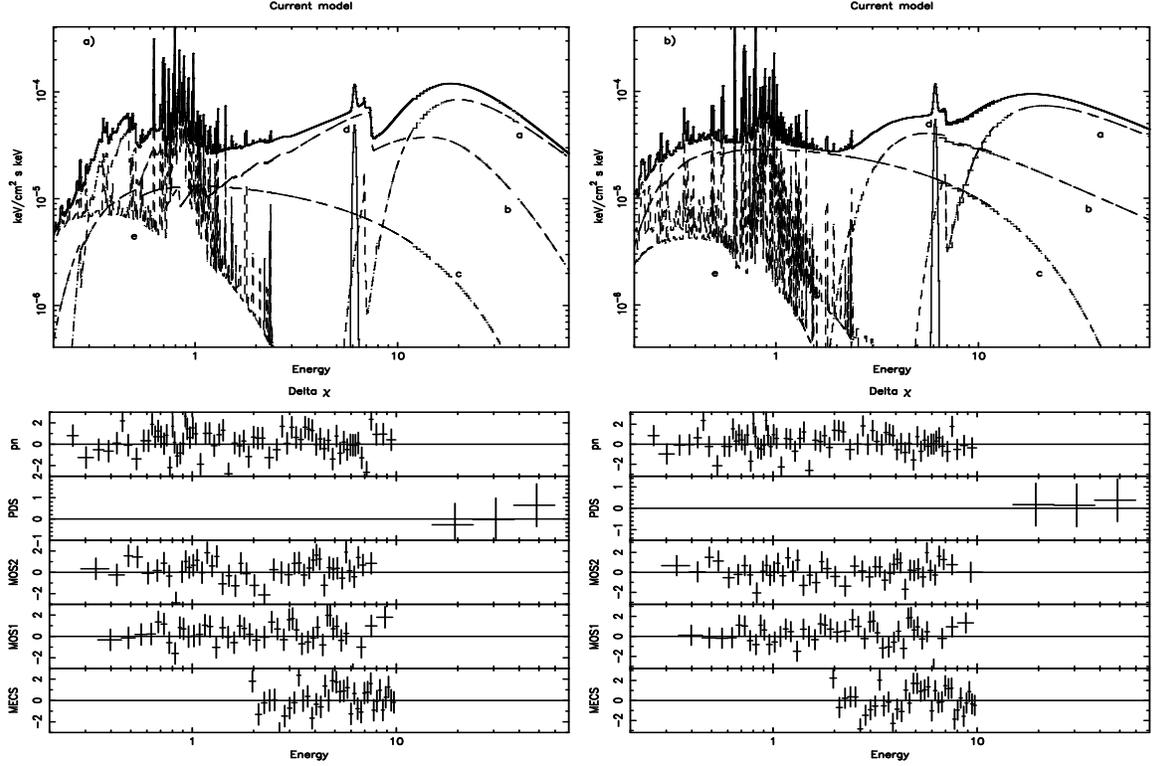

\skip-0.2truecm
\begin{tabular}{cc}
\hskip-0.2truecm\psfig{file=fig2a.ps,height=5.0cm,width=7.5cm,angle=-90}&
\hskip-0.2truecm\psfig{file=fig2b.ps,height=5.0cm,width=7.5cm,angle=-90}\\
\hskip-0.2truecm\psfig{file=fig2c.ps,height=5.0cm,width=7.5cm,angle=-90}&
\hskip-0.2truecm\psfig{file=fig2d.ps,height=5.0cm,width=7.5cm,angle=-90}\\ 
\end{tabular}
 \caption{0.2--50 keV data coverage.
 Upper panels:  best fit models for the
reflection-dominated (left panel) and scattering-dominated scenario (right panel). The spectral components are: a) a highly absorbed
PL AGN;  b) pure reflected AGN  from slightly ionized
material (left panel) or  scattered AGN component (right panel); c) a cutoff PL  associated with the binaries in the
starburst; d) a narrow Gaussian line at 6.39 keV; 
 and e) a thermal emission component associated with the starburst. Lower
panels:  residuals for the different detectors (from up to down: pn, PDS, MOS2, MOS1, MECS).}\label{fig2}
\end{figure*}

The 0.5--50 keV  spectrum of MKN~231 is  particularly complex with both the signatures of the powerful  starburst and of a heavily
obscured AGN; a more
detailed discussion of the spectral modeling   is reported in
\cite{Braito}. \\ 

The 2--10 keV  spectrum   confirms that MKN~231 is characterized  by: a  weak Fe emission line ($EW\sim
300$ eV) and a very
hard  X-ray emission  ($\Gamma\sim0.8$ if a single power law (PL)  model is fitted) with observed  2--10
keV luminosity of $\sim5\times 10^{42}$ erg s$^{-1}$. This latter  is a factor 50 lower than what expected from
the bolometric luminosity.  All these  observational evidences suggest that in the 2--10 keV bandpass we
are seeing only reprocessed  emission (through reflection or scattering on a Compton
thick  mirror).

 The  AGN  emission  (see Fig.~\ref{fig2}) can be
decomposed in:
\begin{itemize}
\item a transmitted  PL which emerges above 10 keV (component a in  Fig.~\ref{fig2})
and is filtered by a  high column density screen 
($N_{H}\sim2\times10^{24}$ cm$^{-2}$);
 \item  a reprocessed (through scattering  or reflection) PL continuum, which dominates the emission in the 2--10 keV
 range (component b). This reprocessed emission is also absorbed ($N_{H}\sim 10^{21}-10^{22}$ cm$^{-2}$);
 \item  a Fe K$\alpha$ emission line ($E_c =6.39\pm0.15$ keV, $EW=290\pm 110$ eV; component d).
\end{itemize}

The  2--10 keV   continuum  and the weak Fe emission line  could  be explained   with reflection  from slightly ionized material
or  assuming  that we see the AGN mainly  through  scattered emission
(see \cite{galla} modeling of the {\it Chandra} data). In the  latter scenario we assume that the Fe
emission line is produced by transmission and is diluted from the scattered component.\\
 
It is worth  noticing that,  regardless of the  possible models assumed  for the hard
X-ray emission (scattering or reflection), we always found   a  soft (0.5--2 keV)
X-ray component   associated with the starburst activity ($L_{(0.5-2)}=6-9 \times 10
^{41}$erg s$^{-1}$). Furthermore in the 
proposed models another PL component is required to account for the 2--10 keV emission
of MKN~231 ($L_{(2-10)}=0.7-1.6 \times 10^{42}$erg s$^{-1}$); this latter component 
could be identified with  the hard X-ray emission
from a population of  High Mass X-ray binaries (component c in 
Fig.~\ref{fig2}), which are expected to be copious in a strong starburst source like MKN 231. 

\section{Main Results}

The main result reported here is that a highly absorbed (N${_H}\sim 2\times 10^{24}$ cm$ ^{-2}$) AGN
component having an intrinsic 2--10 keV luminosity $1^{+1.0}_{-0.5}\times 10^{44}$erg/s has been
detected (the error on this estimate refers to the
range of  luminosities derived with the two  models considered).
{\bf This is the first direct measurement of the intrinsic power of the AGN hosted by MKN~231.}\\ 

This luminosity is about a factor 50 less than the infrared luminosity.
Thus, the total FIR luminosity of the system cannot be entirely associated 
with the AGN, even assuming an AGN UV luminosity a factor 10 greater than the X-ray luminosity
(as observed in QSOs; e.g. \cite{Elvis}).
This suggest that the bulk FIR emission may be due 
to the starburst, in agreement with the results obtained  modeling of the 1--1000 $\mu m$ 
Spectral Energy Distribution \cite{Farrah}.

These observations  give also some suggestions on the physical status of 
the material surrounding the active nucleus in MKN 231.
We identify the absorber  which scatters or reflects the primary
emission, with the BAL outflows which originate close to the central source. In
particular the ionized mirror could be the inner part of  this outflows  (i.e. the
``shielding gas'' proposed by \cite{galla,mur}). In both  models proposed for the AGN
emission, the scattered/reflected components are absorbed.  This latter absorbing medium
could be identified with the  starburst  regions (see e.g. \cite{lev})  or with a
different line of sight through the BAL wind. \\

\end{document}